\documentclass{elsart}
\usepackage[dvips]{graphicx}
\usepackage{amssymb}

\begin{document}

\begin{frontmatter}
\title{Electron-Stimulated Emission of Na Atoms from NaCl Nanocube Corners}
\author[sissa,democritos]{D. Ceresoli},
\author[sissa,democritos]{T. Zykova-Timan},
\author[sissa,democritos,ictp]{E. Tosatti\corauthref{corresponding}}
   \corauth[corresponding]{Corresponding author.}
   \ead{tosatti@sissa.it}
\address[sissa]{International School for Advanced Studies (SISSA),
   via Beirut 2, 34014 Trieste, Italy}
\address[democritos]{Democritos National Simulation Center,
   Trieste, Italy}
\address[ictp]{International Center for Theoretical Physics (ICTP),
   Strada Costiera 11, 34014, Trieste, Italy}

\begin{abstract}
We performed first principles density functional calculations and simulations 
of magic-size neutral NaCl nanocubes, and computed the the extraction of a 
Na neutral corner atom after donating an electron. The atomic  structure 
of the resulting Na corner vacancy is presented.
\end{abstract}

\begin{keyword}
Alkali halides \sep
Nanoclusters \sep
Density Functional Theory \sep
Surface energy \sep
Atom emission
\end{keyword}
\end{frontmatter}

\section{Introduction}\label{sec:intro}
Alkali halide compounds such as NaCl form very stable ionic crystals. Their 
binding energy consists almost entirely of Coulomb attraction of the excess 
electric charges of oppositely charged ions. In this respect, extraction of 
ions, or of neutral molecules, from a bulk alkali halide crystal is a very 
endoenergetic process. For instance, the energy required to remove a NaCl 
molecule from bulk NaCl is 2.25 eV per molecule. Following addition or removal 
of an electron, individual neutral atoms can also be extracted from a surface. 
That process however is expected to be energetically less expensive. 

For the extraction of halogen atoms, the energy scale is set by the exciton 
energy of NaCl -- roughly 8 eV. In fact, once the solid is primed with one 
(surface) exciton it is known to emit a halogen atom with creation of a surface 
F-center~\cite{Fowler}. Recent experiments with STM tips on NaCl(100) thin 
films~\cite{Li92,Meyer}, also showed that it is possible to extract halogen 
atoms under conditions of applied voltage. 

Here we are concerned with the emission of neutral Na atoms upon donation of 
electrons. Experimentally, it is observed that some neutral NaCl nanoclusters
become negative through electron attachment or positive via 
photoemission.~\cite{Bloomfield}
Earlier semi-empirical estimates on NaCl nanoclusters indicates neutral Na 
atom emission with a small activation energy of 0.4 eV~\cite{Diefenbach85,%
Galli86,Martin93}. In nanocube clusters, such low detachment energies can 
arise from the weakly bound corner atoms. Availability of cheap reactive 
alkali atoms might prove of considerable practical importance in the future. 
Motivated by this observation, we addressed the problem of calculating -- 
now from first principles -- the energy necessary to detach a neutral corner 
alkali atom after donation of an electron to an alkali halide crystalline 
nanocube.

Let us begin by considering an infinite surface first. An electron added to 
an infinite NaCl(100) surface will presumably go into a surface state just 
below the the conduction band bottom $\sim$ 1 eV below the vacuum level.
By calculating the electronic structure of a neutral NaCl(100) slab, we find 
the empty surface state of NaCl(100) to be a linear combination of surface Na 
3$s$ and 3$p_z$ orbitals. The added electron could remain delocalized over the 
surface, or else it could lead to a local deformation and become localized 
and self-trapped. If it became self-trapped near a single Na$^+$ surface ion, 
the added electron would neutralize it, thus canceling the coulomb attraction 
that binds the now neutral Na surface atom in its lattice place. As a result, 
this Na atom will remain bound to the surface only by weak induced-dipole and 
dispersion forces, making it relatively easy to detach. 
For this reason, emission of neutral Na atoms after 
electron addition can be expected to happen with a relatively modest energy 
cost. The final outcome after neutral atom emission, will be a surface atom 
vacancy, associated with a large characteristic rearrangement of the 
surrounding lattice ions.

The amount of energy required for extraction of an atom is expected to 
decrease for atoms located in less favorable positions, such as at surface 
steps and kinks, where the local coordination is smaller. The most favorable 
case of all should be represented by corner atoms of alkali halides cubes, 
where the ionic binding is weakest. Moreover, there is in this case no 
question of self-trapping, since the corner atoms naturally trap the carrier 
preferentially over all other atoms.

\section{Computational details}\label{sec:computational}
NaCl nanoclusters were built by cutting out cubelets from a perfect NaCl 
crystal, with initial interatomic Na--Cl distance of 2.82 \AA.

Previous calculations of electron attachment to a NaCl nanocluster exist 
based on a quantum path-integral molecular dynamics method, using classical 
interatomic potentials, and treating the quantum nature of the extra electron 
in an effective way~\cite{Landman85,Landman95}.

We performed total energy calculations of NaCl(100) and of cubic NaCl
nanoclusters in the first principles framework of density functional theory 
(DFT). We used the plane wave pseudopotential method~\cite{PWSCF,CPMD},
with norm-conserving pseudopotentials~\cite{MT} and a plane wave cutoff of
40 Ry. Nonlinear core corrections~\cite{NLCC} were included in the
pseudopotential for Na. All calculations were spin polarized and the 
gradient-corrected BLYP exchange-correlation functional~\cite{BLYP} was used.

Initially electron selfconsistency and full zero-pressure equilibrium was 
generated as a starting point of the neutral cluster. Donation of an electron 
was simulated by adding an electron in the LUMO, and by neutralizing the 
system by a uniform compensating background. The total energy change resulting 
by adiabatically pulling the appropriate surface or corner Na atom away from 
its initial location gives the desired extraction energy. We note that the 
energetics of this system may in principle be affected by finite-size
effects and by self-interaction if the extra electron is localized. 
The finite size effects can be checked by comparing results for increasing
size. The self-interaction error will only be large if the localization 
is extreme, and can still be be minimized when dealing with initial and final 
adiabatic states with similar levels of electron localization. 
In our case, we explicitly checked the ionization energy of an isolated 
Na atom and found it to be 5.3 eV, which is close to the experimental 
value 5.139 eV indicating that self-interaction is not an important problem.

\section{Results and discussion}\label{results}
It is known from experiment~\cite{Martin93} that ionized NaCl nanoclusters
have ``magic numbers'' corresponding to small cubelets exposing (100) facets,
whose general formula is [Na$_n$Cl$_{n-1}$]$^{+}$ or [Na$_n$Cl$_{n+1}$]$^{-}$.
Even if neutral clusters cannot be detected in mass-spectroscopy experiments,
they are also predicted~\cite{Martin93} to be of the cubic shape, with general
formula Na$_{2n}$Cl$_{2n}$. Charged clusters have an odd number of ions on 
the cubic edges, whereas neutral clusters have an even number of ions.
Positively charged clusters, in this idealized picture, have Na$^+$ corner 
ions. Negatively charged clusters have Cl$^-$ corner ions. On the contrary, 
neutral clusters will have four Na$^+$ and four Cl$^-$ corner ions.

\begin{figure}\begin{center}
   \includegraphics[width=0.8\textwidth]{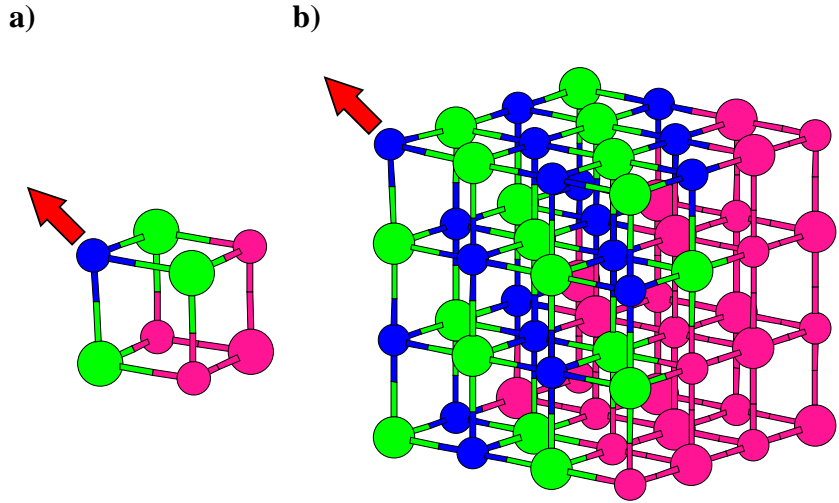}
   \caption{(color online) a) small cube Na$_4$Cl$_4$.
   b) large cube Na$_{32}$Cl$_{32}$. Small circles: sodium; Large circles: 
   chlorine; purple circles: fixed atoms (both Na and Cl). The red arrow 
   indicates the direction of extraction of the Na corner atom.}
   \label{fig:nanocubi}
\end{center}\end{figure}

We studied neutral nanocubes of two sizes: the small one just made up of
8 atoms or 4 molecules (Na$_4$Cl$_4$); the large one of 64 atoms or 32
molecules (Na$_{32}$Cl$_{32}$). The nanocubes were periodically repeated in
a cubic box of 10 \AA\ and 20 \AA\ of side, respectively, where the residual 
interaction with periodic replicas is negligible. One Na atom was displaced 
gradually from one corner in the (111) direction of the cubic supercell. 
The displacement was measured from the ideal corner coordinates of the 
perfect cube, taken as the reference geometry. The geometry was relaxed for 
outwards displacements of the extracted corner atom of 0.1 \AA, 0.2 \AA\ and 
5.0 \AA. In order to simulate the mechanical rigidity of a much bigger cube, 
only ions closest to the resulting corner vacancy were allowed to relax, 
whereas the remaining nanocube ions (roughly 1/2 of the total) were kept 
fixed as indicated by different coloring in Fig.~\ref{fig:nanocubi}.
$\Gamma$-point sampling was used in all calculations.

The total energy was found to increase monotonically as a function of 
displacement, leveling off for an outward displacement of $\sim$ 4 \AA\ and 
higher. The extraction energy is evaluated as a difference between the total 
energy for a displacement of 5.0 \AA\ and the total energy for zero 
displacement.

As anticipated, initially the energy for zero displacement may not be directly 
comparable to the final distorted energy because of the different 
self-interaction energy. For example one added electron would in the perfect 
cubic cluster be perfectly delocalized, because of symmetry, over the four Na 
corner atoms. Due to this in the initial (delocalized) state the electron 
self-interaction is much lower than in the distorted (localized) state, and 
the total energy difference may be affected. However as soon as one corner Na 
atom is even slightly displaced outwards, the four-fold degeneracy is removed 
and the extra electron wave function strongly localizes on that Na atom.
In this manner the error is approximately eliminated. The 
effective initial state energy is therefore calculated by assuming a very 
small displacement breaking the cubic symmetry in the initial state too,
and then extrapolating to zero the initial displacement. 
For an appreciation of the effect mentioned above, the perfect cubic 
Na$_{32}$Cl$_{32}$ cluster energy is $E_0=-$~1687.173~eV, that with a slight 
distortion (extrapolated to zero) is $E_0'=-$~1687.211~eV, and the final one 
after full Na atom extraction $E_5'=-$~1686.531~eV.
As one can see, the difference $E_0 - E_0'=$~38~meV (which also contains the 
self-interaction), is negligibly smaller compared  the Na extraction energy
$\Delta = E_5'-E_0' =$~0.68~eV.

\begin{table}\begin{center}
\begin{tabular}{cccc}
   \hline\hline
    & small cube (Na$_4$Cl$_4$) & large cube (Na$_{32}$Cl$_{32}$) \\
   \hline
   Na extraction energy $\Delta$ & 0.62 eV & 0.68 eV \\
   (electron addition) & & & \\
   \hline\hline
\end{tabular}
\caption{Energy necessary to extract a Na neutral atom from the
corner of a nanocube.} 
\label{tab:energies}
\end{center}\end{table}

\begin{figure}\begin{center}
   \includegraphics[width=0.4\textwidth]{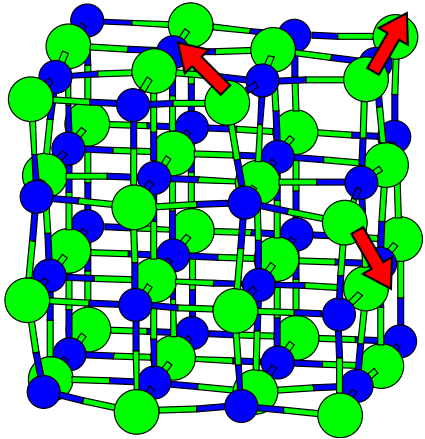}
   \caption{(color online) Relaxation pattern around the Na corner
   vacancy (top right) in the large cube Na$_{32}$Cl$_{32}$. 
   Small circles: Na(+); Large circles: Cl(-).}
   \label{fig:relaxation}
\end{center}\end{figure}

Our preliminary calculated energies for removal of neutral corner Na atoms 
are shown in Tab.~\ref{tab:energies} for the small and large cubes.
The near coincidence of results of the small and large clusters indicates
that finite-size errors arising from various sources, including interactions
between replicas, are relatively unimportant.
The energy required to extract a neutral Na atom, by addition of an electron
is calculated to be $\sim$ 0.6 eV, in fairly good agreement with the 
previous estimates of 0.4 eV~\cite{Martin93,Diefenbach85,Galli86}
The relaxation pattern near the corner defect is shown in 
Fig.~\ref{fig:relaxation}: the three first neighbors Cl$^-$ ions relaxed 
outwards in such a way as to minimize their repulsion. The extra electron 
goes into the Na 3$s$ level, which falls in the middle of the NaCl cluster
energy gap.

\section{Conclusion}\label{conclusion}
We calculated the extraction preocess of a neutral Na corner atom from NaCl
nanoclusters to which one electron has been donated. 
A relatively low extraction energy of the order of half an eV previously
suggested~\cite{Diefenbach85,Galli86,Martin93} is essentially confirmed by 
the calculation. This implies that neutral Na atoms will be to some extent 
thermally emitted by corner sites of negatively charged NaCl nanocubes. In 
conclusion the nanocube corners seem to be an  energetically cheap source 
of very reactive Na atoms.

\section*{Acknowledgments}
Project sponsored by Italian Ministry of University and Research, through
MIUR Cofin 2003028141 and Cofin 2004023199, and by FIRB RBAU01LX5H and 
FIRB RBAU017S8R; and by INFM, through ``Iniziativa Trasversare Calcolo 
Parallelo''. Calculations were performed on the SP5 and CLX clusters at 
CINECA, Casalecchio (Bologna). We acknowledge discussion with F. Stellacci 
and private communications by G. Meyer.




\begin{thebibliography}{00}
\bibitem{Fowler} 
  M. Georgiev, \emph{F--centers in alkali halides}, Springer, Berlin, 1988. 

\bibitem{Li92}
   X. Li, R. D. Beck and R. L. Whetten,
   Phys.\ Rev.\ Lett.\ \textbf{68}, 3420 (1992).

\bibitem{Meyer}
  J. Repp, G. Meyer, S. Paavilainen, F. E. Olsson and M. Persson,
  private communication 

\bibitem{Bloomfield}
  C. W. S. Conover, Y. A. Yang, and L. A. Bloomfield
  Phys.\ Rev.\ B \textbf{38}, 3517 (1988).

\bibitem{Diefenbach85}
  J. Diefenbach and T. P. Martin, J.\ Chem.\ Phys.\ \textbf{83}, 4585 (1985).

\bibitem{Martin93}
   T. P. Martin, Phys.\ Repts.\ \textbf{95}, 168 (1983).

\bibitem{Galli86}
  G. Galli, W. Andreoni and M. P. Tosi, Phys.\ Rev.\ A \textbf{34}, 3580 (1986).

\bibitem{Landman85}
   U. Landman, D. Scharf and J. Jortner,
   Phys.\ Rev.\ Lett.\ \textbf{54}, 1860 (1985).

\bibitem{Landman95}
   R. N. Barnett, H.-P. Cheng, H. Hakkinen and U. Landman,
   J.\ Phys.\ Chem.\ \textbf{99}, 7731 (1995).

\bibitem{PWSCF}
   S. Baroni, A. Dal Corso, S. de Gironcoli, P. Giannozzi, C. Cavazzoni,
   G. Ballabio, S. Scandolo, G. Chiarotti, P. Focher, A. Pasquarello,
   K. Laasonen, A. Trave, R. Car, N. Marzari, A. Kokalj,
   \textsl{http://www.pwscf.org}

\bibitem{CPMD}
   CPMD, Copyright IBM Corp 1990-2004, Copyright MPI f\"ur
   Festk\"orperforschung Stuttgart 1997-2001.

\bibitem{MT}
   N. Troullier and J. L. Martins, Phys.\ Rev.\ B \textbf{43}, 1993.

\bibitem{NLCC}
   S. G. Louie, S. Froyen and M. L. Cohen,
   Phys.\ Rev.\ B \textbf{26}, 1738 (1982).

\bibitem{BLYP}
   A. Becke, Phys.\ Rev.\ A \textbf{37}, 785 (1988);
   C. Lee, W. Yang and R. Parr, Phys.\ Rev.\ B \textbf{37}, 785 (1988).


\end{thebibliography}
\end{document}